# Ensuring Adherence to Standards in Experiment-Related Metadata Entered Via Spreadsheets

Martin J. O'Connor[1*], Josef Hardi[1*], Marcos Martínez-Romero[1], Sowmya Somasundaram[1], Brendan Honick[2], Stephen A. Fisher[3], Ajay Pillai[4], Mark A. Musen[1]

[1]Stanford Center for Biomedical Informatics Research
Stanford, CA 94305, USA
[2]Pittsburgh Supercomputer Center
Pittsburgh, PA 15213, USA
[3]University of Pennsylvania
Philadelphia, PA 19104, USA
[4]National Human Genome Research Institute
Bethesda, MD 208792, USA
[*]These authors contributed equally to this work.

## Abstract
Scientists increasingly recognize the importance of providing rich, standards-adherent metadata to describe their experimental results. Despite the availability of sophisticated tools to assist in the process of data annotation, investigators generally seem to prefer to use spreadsheets when supplying metadata, despite the limitations of spreadsheets in ensuring metadata consistency and compliance with formal specifications. In this paper, we describe an end-to-end approach that supports spreadsheet-based entry of metadata, while ensuring rigorous adherence to community-based metadata standards and providing quality control. Our methods employ several key components, including customizable templates that represent metadata standards and that can inform the spreadsheets that investigators use to author metadata, controlled terminologies and ontologies for defining metadata values that can be accessed directly from a spreadsheet, and an interactive Web-based tool that allows users to rapidly identify and fix errors in their spreadsheet-based metadata. We demonstrate how this approach is being deployed in a biomedical consortium known as HuBMAP to define and collect metadata about a wide range of biological assays.

## Introduction
Metadata, which provide descriptive information about data, are an essential component of the scientific endeavor. Accurate and comprehensive metadata that adhere to community standards are essential if datasets are to be findable, accessible, interoperable, and reusable (FAIR) (1) and if datasets are to be understood, analyzed, and reused by other researchers. Many scientific communities have created standardized reporting guidelines that specify the structure of the domain-specific and experiment-specific metadata required to meet these goals, enumerating the attributes of the experimental situation and of the data that the metadata need to describe. For example, the Minimal Information About a Microarray Experiment (MIAME) standard (2) enumerates the attributes of experiments in functional genomics that should appear in all microarray-related metadata, and it is among the best known reporting guidelines in science.

Metadata overall can be seen as a list of attribute–value pairs. To ensure high levels of standardization, controlled terminologies and ontologies commonly provide constraints on the values allowed for many of the attributes in the list.

While metadata reporting guidelines provide a core structure for high-quality metadata, there are many practical impediments when researchers attempt to author metadata that conform to such guidelines. A key challenge is providing capabilities for authoring rich, standards-adherent metadata that fit with existing laboratory workflows and computer-based tools. In particular, solutions that interoperate seamlessly with spreadsheets—overwhelmingly the most popular data-entry instrument for researchers—can help address these challenges, especially when multiple assays are performed in batch mode, resulting in large collections of related metadata records.

Scientists use spreadsheets to acquire data and metadata, to exchange these data and metadata with collaborators, and to analyze their results in tools such as Excel and Google Sheets. While spreadsheets are powerful tools for data management, they are not good at enforcing adherence to community standards. Adherence errors can include missing required fields, typos, formatting errors, or values that do not conform to pre-specified value sets indicated in the metadata specification. While some validation is possible within spreadsheet tools, it is quite limited. Tools such as Excel, for example, do allow users to set rules for what type of data is allowed in a cell or in a range of cells. Excel dropdowns can also be used to suggest a set of pre-defined values for some cells, which can encourage users to supply correct values from pre-specified lists.

While these features encourage adherence to metadata specifications, they cannot ensure it. Users are still free to blithely ignore the constraints and to supply erroneous metadata. As a result, metadata ingestion processes must anticipate and handle a large variety of possible errors in user-submitted metadata spreadsheets. Fixing many of these errors can require manual intervention from curators, which may necessitate contacting submitters to help rectify those errors—assuming that the errors are obvious. The end result can be a poor user experience for metadata submitters and expensive-to-maintain ingestion pipelines, all of which can lead to poor quality metadata and to a lack of data FAIRness.

There is a pressing need for tools that can facilitate the creation of high-quality, standards-adherent metadata and that can still allow researchers to use their familiar spreadsheets. In this paper, we outline such a system. Building on an existing metadata management platform known as the CEDAR Workbench (3), we outline how we have developed an end-to-end solution for encoding metadata specifications, for representing those specifications as spreadsheets, and then for ensuring strong compliance to those specifications when metadata are acquired. Our approach employs customizable templates for defining metadata reporting guidelines, integral support for the use of controlled terminologies and ontologies, and an interactive Web-based tool for metadata validation and repair. We show how the resulting system provides metadata specification, acquisition, validation, and repair capabilities that can help ensure high-quality metadata. We demonstrate how our approach is being deployed in a large biomedical consortium to define and collect metadata about scientific experiments, and how it provides an effective and efficient solution for ensuring high-quality metadata in spreadsheet-based metadata-acquisition systems.

Several tools have been developed to address the limitations of spreadsheet-based metadata acquisition systems.

One of the earliest of these tools is RightField (4). RightField is designed to improve the quality of spreadsheet-based data by guiding users to provide consistent metadata values during data entry. It allows researchers to restrict cells to contain standardized terms from ontologies and other controlled terminologies in the BioPortal ontology repository (5). RightField templates can include dropdown lists and other validation rules to ensure the data entered are consistent. RightField also allows users to export collected data in various formats, including the ISA-Tab format (6), which is a framework for describing and sharing certain kinds of metadata and experimental results in life-sciences research.

OntoMaton (7), which is also designed to work with the ISA-Tab format, is a similar tool that supports the enforcement of the use of standard terminologies in spreadsheets. The tool allows users to search and access data from various biomedical ontologies that are present in BioPortal directly from within Google Sheets. It provides a user-friendly interface that allows users to search for terms within these ontologies and it automatically retrieves related terms and associated metadata, such as definitions and synonyms.

Mapping Master (8) adopts an approach different from that of these other tools. It uses spreadsheets as a starting point and provides mechanisms to map their content to OWL ontologies. It thus allows users to create OWL ontologies using spreadsheet content (9). For example, users could use it to create a set of OWL classes named using the content of each cell in a particular column. Mapping Master uses a drag-and-drop interface to make the mapping process easy and intuitive. By mapping data to standard vocabularies, Mapping Master can quickly identify errors in spreadsheets and help to ensure that metadata are well-defined and interoperable.

SWATE (Swate Workflow Annotation Tool for Excel) (10) is a spreadsheet-based tool developed to facilitate ontology-driven metadata annotation, particularly for workflows in plant research. Integrated directly into Microsoft Excel, SWATE allows researchers to annotate experimental data using controlled vocabularies from ontologies stored in an internal database. The tool supports a range of features, including pre-defined and customizable templates aligned with standards like MIAPPE (Minimum Information About a Plant Phenotyping Experiment).

Sasse et al. (11) provide a comprehensive review of semantic metadata annotation tools in the biomedical domain, evaluating their alignment with FAIR principles and their ability to enhance data interoperability. The study identifies key features required for effective metadata annotation, such as integration with existing workflows and support for retrospective semantic enrichment.

In addition to annotating, some tools concentrate on providing mechanisms to repair spreadsheet-based data. One of the most popular of these types of tools is OpenRefine (12), which provides an interface that allows users to easily clean, transform, and organize their data. With OpenRefine, users can split, merge, and reorder columns, as well as filter and manipulate rows based on various criteria. It also provides advanced features for detecting and correcting errors, removing duplicates, and transforming data into different formats.

None of these tools enforces adherence to discipline-specific metadata reporting guidelines in a flexible manner. Either a rather generic metadata structure (i.e., ISA) or a specific domain are assumed, or the attributes of the metadata associated with a particular guideline need to be entered into the spreadsheet by hand. Our approach puts emphasis on the reporting guideline for the metadata as a first-class entity, and it ensures that user entries comport with whatever guideline is relevant for the particular class of experiment for which metadata need to be entered.

Some expertise is required when using these other tools, and setting up a structured spreadsheet for metadata entry may require a steep learning curve. In many situations, approaches that can be used by non-specialists for ensuring quality in spreadsheet-based content are desirable. Ideally, these interfaces should be able to quickly identify errors and to suggest repairs. An additional goal includes support for comma- and tab-separated files (CSVs and TSVs, respectively), since these remain common formats in many scientific domains. The ultimate goal is to make human curation more efficient by reducing the need for manual intervention in messy metadata files.

In this paper, we describe technologies we have developed to tackle the problem of standards-based metadata entry using spreadsheets. We show how these technologies were implemented and deployed in the Human BioMolecular Atlas Program (HuBMAP) to support the creation and submission of spreadsheet-based metadata at scale.

## Materials and Methods

The technologies described in this paper were driven by the needs of HuBMAP, a research initiative that aims to identify biomarkers that distinguish every cell in the human body with the goal of creating a multi-scale spatial atlas of the healthy human body at single-cell resolution (13). HuBMAP investigators perform a wide range of assays on many different tissues, with the goal of mapping the body at single-cell resolution. The needs to manage metadata and data describing thousands of biological assays, to ensure that metadata are adherent to standards and that the data are FAIR, and to accommodate the desire of investigators to stick with their familiar spreadsheet-based metadata-entry methods drove us to create extensions to an existing metadata management system that we developed known as the CEDAR Workbench (14).

AutoSave RNAseq HuBMAP Metadata — Saved

Home Insert Draw Page Layout Formulas Data Review View Automate

| | A | B | C | D | E | F |
|---|---|---|---|---|---|---|
| 1 | parent_sample_id | lab_id | preparation_protocol_doi | dataset_type | analyte_class | is_targe |
| 2 | HBM978.QPFT.528 | 3252_ftL_RNA_T1 | https://dx.doi.org/10.17504/protocols.io.4r3l224p3l1y/v1 | RNAseq | RNA | No |
| 3 | HBM549.HWXD.539 | 3252_ftR_RNA_F1 | https://dx.doi.org/10.17504/protocols.io.4r3l224p3l1y/v1 | RNAseq | RNA | No |
| 4 | HBM263.RDXS.575 | 3252_um_RNA_E1 | https://dx.doi.org/10.17504/protocols.io.4r3l224p3l1y/v1 | RNAseq | RNA | No |
| 5 | HBM463.PFCR.978 | 3252_um_RNA_S1 | https://dx.doi.org/10.17504/protocols.io.4r3l224p3l1y/v1 | RNAseq | RNA | No |
| 6 | HBM658.JHJR.554 | 3257_ftL_RNA_F1 | https://dx.doi.org/10.17504/protocols.io.4r3l224p3l1y/v1 | RNAseq | RNA | No |
| 7 | HBM538.BXDM.784 | 3257_ftL_RNA_T1 | https://dx.doi.org/10.17504/protocols.io.4r3l224p3l1y/v1 | RNAseq | RNA | No |
| 8 | HBM553.MJZT.253 | 3257_um_RNA_E1 | https://dx.doi.org/10.17504/protocols.io.4r3l224p3l1y/v1 | RNAseq | RNA | No |
| 9 | HBM794.LJLS.323 | 3257_um_RNA_S1 | https://dx.doi.org/10.17504/protocols.io.4r3l224p3l1y/v1 | RNAseq | RNA | No |
| 10 | HBM347.LXXJ.548 | 3276_ftL_RNA_F1 | https://dx.doi.org/10.17504/protocols.io.4r3l224p3l1y/v1 | RNAseq | RNA | No |
| 11 | HBM867.PSHH.452 | 3276_ftL_RNA_T1 | https://dx.doi.org/10.17504/protocols.io.4r3l224p3l1y/v1 | RNAseq | RNA | No |
| 12 | HBM269.TSPN.869 | 3276_um_RNA_E1 | https://dx.doi.org/10.17504/protocols.io.4r3l224p3l1y/v1 | RNAseq | RNA | No |
| 13 | HBM554.MMQJ.822 | 3276_um_RNA_S1 | https://dx.doi.org/10.17504/protocols.io.4r3l224p3l1y/v1 | RNAseq | RNA | No |
| 14 | HBM374.LHQR.298 | 3322_ftR_RNA_F1 | https://dx.doi.org/10.17504/protocols.io.4r3l224p3l1y/v1 | RNAseq | RNA | No |
| 15 | HBM889.GTGK.573 | 3322_ftR_RNA_T1 | https://dx.doi.org/10.17504/protocols.io.4r3l224p3l1y/v1 | RNAseq | RNA | No |
| 16 | HBM588.GCWV.564 | 3322_um_RNA_E1 | https://dx.doi.org/10.17504/protocols.io.4r3l224p3l1y/v1 | RNAseq | RNA | No |
| 17 | HBM592.ZCTZ.995 | 3322_um_RNA_S1 | https://dx.doi.org/10.17504/protocols.io.4r3l224p3l1y/v1 | RNAseq | RNA | No |

**Figure 1:** Excel-based metadata describing a collection of RNAseq-based data sets. Each row contains metadata for a different data set. Submitters upload these metadata and associated data sets to the HuBMAP data coordinating center.

## Background: HuBMAP

The goal of the HuBMAP Consortium is to accelerate the development of tools and techniques for constructing high-resolution spatial tissue maps of the human body and to establish an open data platform for sharing this knowledge. A key focus of HuBMAP is to make data FAIR by requiring that all submitted metadata adhere to standards developed or endorsed by the HuBMAP community. The consortium has created a robust data-ingestion pipeline to help ensure standards adherence. Using this pipeline, users can submit spreadsheet-based metadata along with associated experimental results.

The HuBMAP Data Coordination Working Group has developed several dozen metadata reporting guidelines, with the majority of these guidelines targeting metadata for a particular type of biological assay. These reporting guidelines include lists of standard attributes of metadata for a variety of experimental assay types. Each specification of a metadata attribute indicates such things as the datatype of the field (e.g., numeric, Boolean, string) and constraints on those values (e.g., allowed value ranges for numeric fields). Sets of allowed values have also been developed for many metadata fields.  Before the implementation of our automated approach, HuBMAP investigators used spreadsheets to acquire instances of metadata compatible with the reporting guidelines, where each field in a guideline became a column header in the spreadsheet, and each row in the spreadsheet represented the metadata specification for a particular experiment (Figure 1).

Before the development of the tools described in this paper, HuBMAP data providers would select the appropriate spreadsheet specific to the experimental assay the used and then populate the spreadsheet with their metadata, filling in each row with the metadata for a different experimental run. They would then upload the metadata along with their data to the HuBMAP data coordinating center. The ingestion pipeline analyzed the submitted metadata for compliance with the metadata specification to the degree possible. If errors were identified in the submission, the ingestion process was halted and an error report was presented to the data submitter. The submitter could then fix those errors and resubmit. While this process eventually converged to an acceptable submission, a significant number of iterations were often needed, and many metadata errors evaded detection. In a considerable number of cases, manual intervention was needed to help submitters supply the correct metadata.

There was thus a pressing need for a solution that allowed users to easily identify and repair errors in their metadata to ensure strong enforcement of metadata quality, while continuing to support metadata submission using spreadsheets. Driven by these needs, we developed a series of extensions to the CEDAR metadata management system. Our solution provides an end-to-end approach for creating, managing, acquiring, validating, and submitting standards-adherent spreadsheet-based metadata.

## Background: The CEDAR Workbench

The Center for Expanded Data Annotation and Retrieval (CEDAR) was established in 2014 to create a computational ecosystem for the development, evaluation, use, and refinement of standards-based, scientific metadata (3). The resulting CEDAR Workbench (12) supports a workflow for metadata management that is organized around three main stages of the metadata development process: (1) representing community-based metadata standards, (2) acquisition of metadata instances comporting with those standards, and (3) submission of metadata to data repositories or data coordinating centers. A driving goal of CEDAR is to provide highly configurable tools that support the creation of metadata-submission pipelines to meet the requirements of a wide range of deployment scenarios. The CEDAR Workbench is a modular system that provides components that can be integrated into existing workflows to address specific tasks in a metadata submission pipeline or that themselves can be assembled to provide an end-to-end pipeline. The system is built around the notion of creating *templates* that represent the structure and semantics of metadata reporting guidelines (15). These templates support a metadata-submission workflow that acquires conforming instances of metadata via custom-generated Web forms and that uploads the resulting metadata to designated repositories (16).

CEDAR's overall metadata workflow comprises the following three steps: (1) Template authors use a CEDAR tool called the Template Designer to create machine-actionable templates that represent metadata reporting guidelines, typically following textual, discipline-specific standards that describe the essential attributes of the types of experiments under consideration (Figure 2).

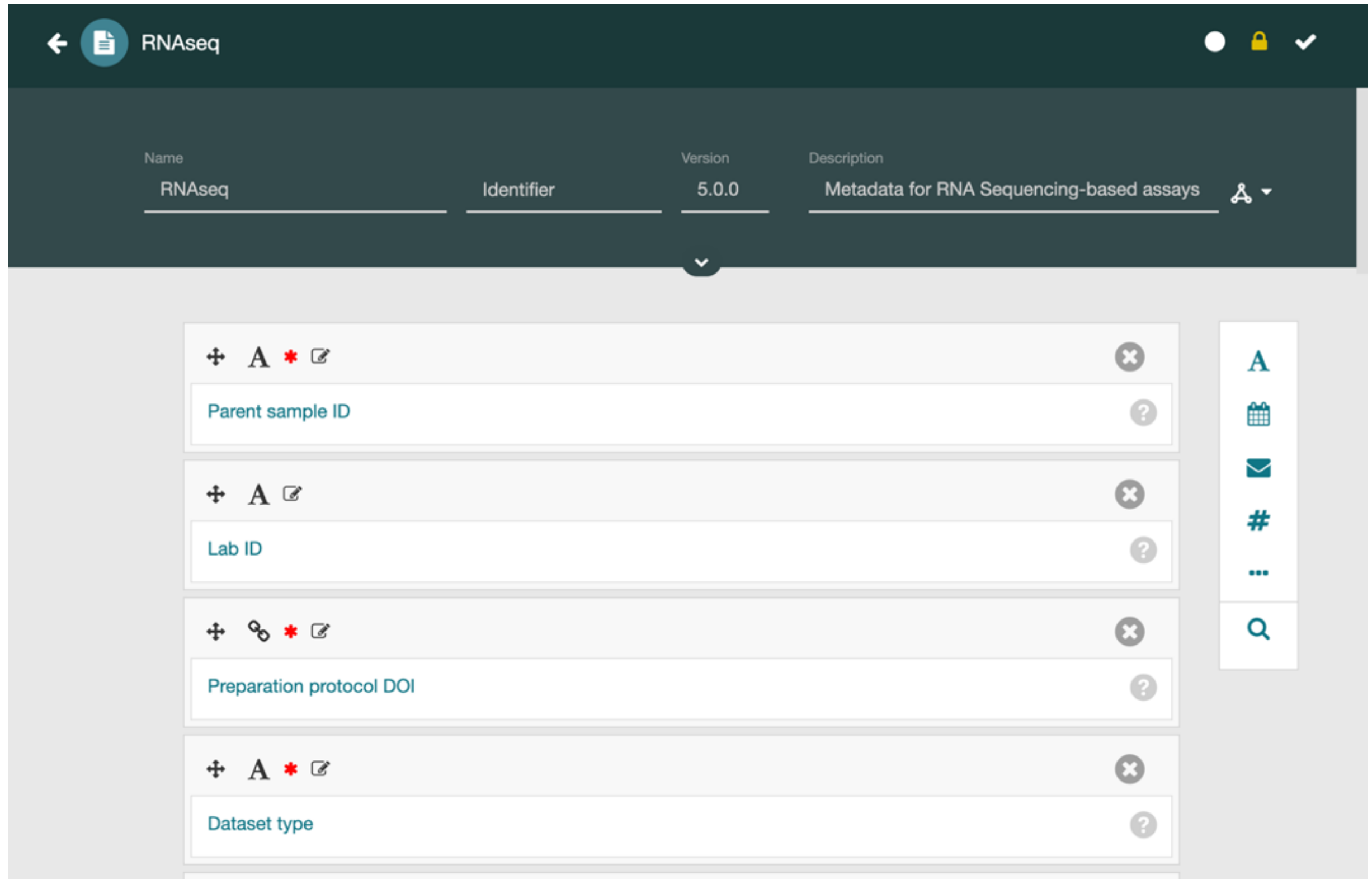

**Figure 2:** A CEDAR template in the Template Designer tool. Template authors can use the Template Designer to interactively build templates to describe metadata reporting guidelines. Here, the template describes metadata for an RNAseq-based assay.

Authors can define their templates to incorporate controlled terms, ontologies, and value sets supplied by the BioPortal ontology repository (7). (2) When a curator or other metadata provider chooses to populate a template, a CEDAR tool called the Metadata Editor automatically generates a form-based user interface from the template (Figure 3); the curator then uses the Metadata Editor to enter the descriptive metadata, creating a metadata instance that conforms with the standard represented by the metadata template. When users populate these metadata-entry forms, semantic information stored in the associated template is used to ensure that metadata entries adhere to the required data types, presenting ontology-controlled suggestions to users and ensuring that the collected metadata conform to the published specification. (3) Once the metadata have been entered, scientists can use the CEDAR Submission Service to upload metadata and associated experimental data to a target repository.

## Extending CEDAR to Support Spreadsheets

We used CEDAR's support for metadata templates that represent community standards to provide a robust, end-to-end spreadsheet-based metadata management solution for HuBMAP.

The first step was to render HuBMAP's metadata reporting guidelines as CEDAR templates.

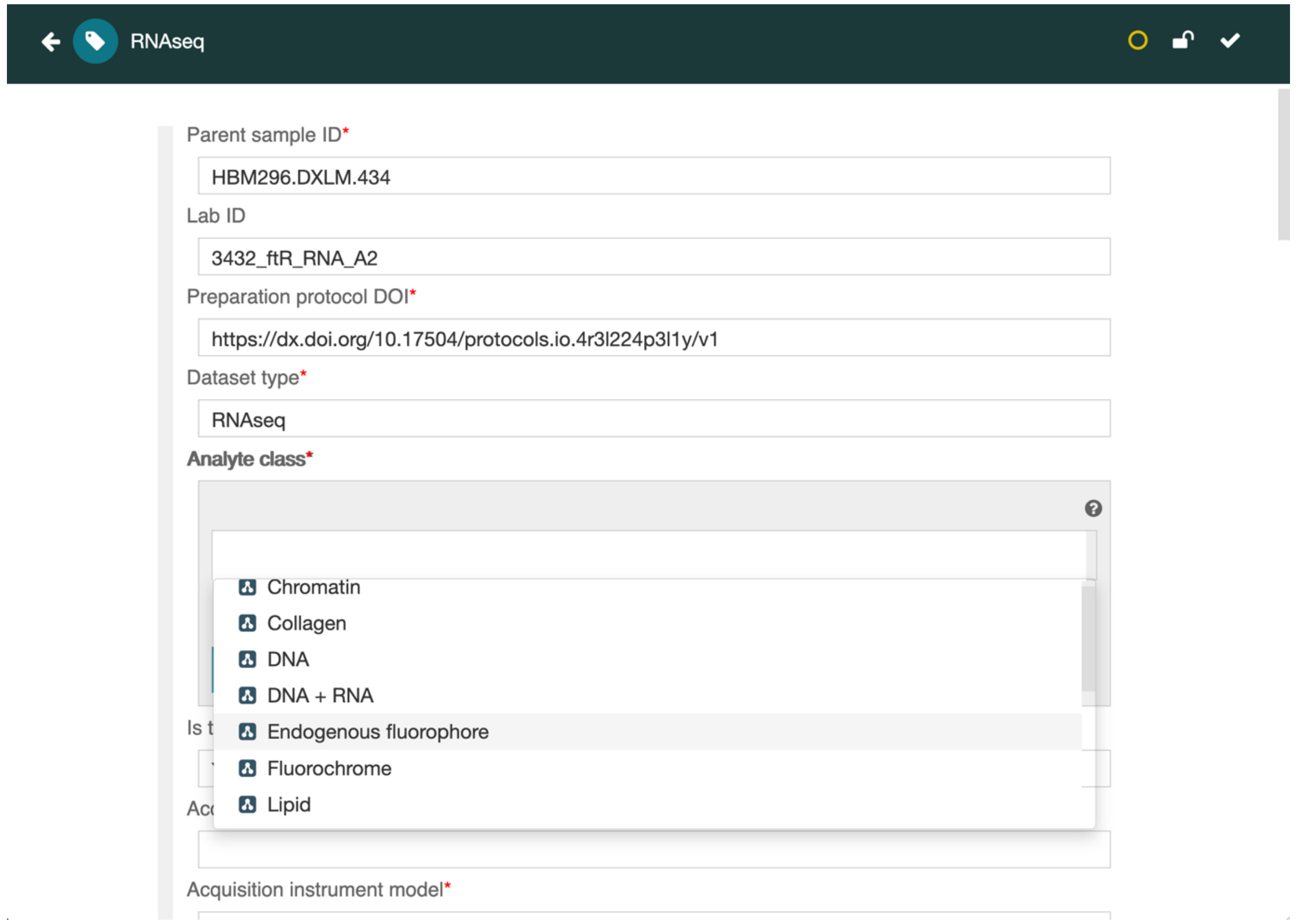

**Figure 3:** CEDAR-generated Web-based form illustrating how metadata can be acquired from users in the Metadata Editor tool. Here, the form is generated automatically from a template that specifies metadata for a RNAseq-based assay. In this case, the metadata author has completed four initial fields and is being presented with a set of choices for a field called *Analyte class*. Once completed, the user can save the completed form, whereupon it is stored in CEDAR.

HuBMAP curators used CEDAR's Template Designer to collaboratively develop new metadata specifications that represent the existing HuBMAP metadata guidelines as machine-actionable templates. The curators made a significant effort to expand the use of controlled terminologies in the templates to increase the quality of the resulting metadata. The use of the CEDAR Template Designer thus made the HuBMAP metadata guidelines more formal and reproducible, and it provided clear, standardized specifications for the data types expected of each metadata attribute.

Although CEDAR's Metadata Editor could provide Web-based metadata-acquisition interfaces for each new HuBMAP metadata template, most HuBMAP consortium members are wedded to metadata-creation processes that are built around spreadsheets. In some cases, these metadata are produced from laboratory instruments as part of computational pipelines and captured directly in spreadsheets. We therefore developed a mechanism to generate spreadsheets that represent metadata reporting guidelines directly from CEDAR templates.

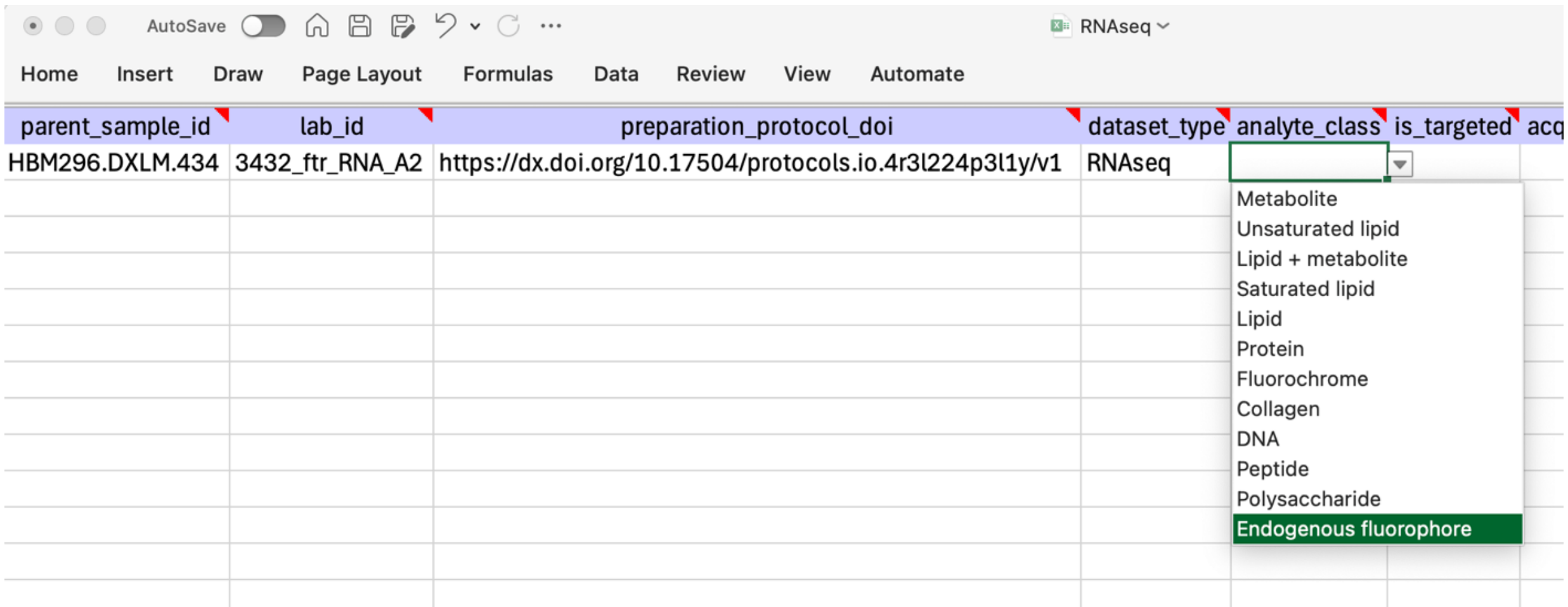

**Figure 4.** CEDAR-generated Excel-based metadata template, illustrating how metadata can be acquired from HuBMAP dataset submitters via spreadsheets. The submitter will first have obtained a copy of a spreadsheet that has been generated from a CEDAR template representing the standard for which they want to supply metadata. This generated spreadsheet will contain column headers for each metadata field in the source template. It will contain constraints on each column to ensure that supplied values are of the correct type. The example shown here was generated from an RNAseq assay template. The user has populated four initial cells and is being presented with a set of choices for a column called *analyte_class*. Since this column contains values restricted to a particular set of values, a set of choices is presented to the user via a drop-down menu. The spreadsheet was generated from the same CEDAR template that was used to generate the Web form shown in Figure 3. Every dataset uploaded to HuBMAP receives a DOI that uniquely identifies the dataset and links to both the data and metadata stored on the HuBMAP site. The DOI for this submission is https://doi.org/10.35079/HBM853.CZKK.264 (17).

As with the initial HuBMAP approach, these spreadsheets offer a structure whereby the *columns* specify the metadata fields required for a particular community-based metadata standard (Figure 4). The spreadsheet to collect metadata for a given class of experiment is created programmatically by CEDAR directly from a source metadata template—just as CEDAR might use the same template to generate a Web form for the Metadata Editor. The spreadsheet-generation process produces Excel-based spreadsheets that aim to constrain the values in each of each column to conform to the specification for corresponding field in the source template. These spreadsheets can then be populated with metadata by end users, such that each *row* contains the metadata for a particular experiment.

When populating these generated spreadsheets, metadata submitters are presented with errors if the values they supply are outside the specified ranges. For example, if a template field is specified to be numeric, the corresponding column in the Excel spreadsheet is constrained to be numeric. This constraint also reflects the datatype of the number, distinguishing between integer and floating-point numbers. If the template specification restricts numbers to designated ranges, these constraints are also reflected in the generated spreadsheet. Similarly, minimum and maximum length constraints are produced for text-based columns. Temporal fields are also handled. Excel allows detailed temporal constraints to be created for columns, allowing, for example, values to be restricted to specified temporal granularities; formatting options are also supported. The spreadsheet generation process uses these constraints to reflect the original temporal type and the formatting options of temporal fields in source templates.

CEDAR also embeds within each spreadsheet—hidden from the user—the value sets and the ontology terms that metadata authors will need to specify standard values for those metadata fields that take on categorical values; these values automatically appear in drop-down menus when users fill in the corresponding spreadsheet cells. Excel's data validation capabilities are used to restrict the values of the corresponding columns to the specified categorical values. In particular, hidden Excel sheets are generated to store the categorical values for each column. The values of each column are then restricted to values contained in its corresponding sheet. The spreadsheet generation process also embeds information in each spreadsheet that indicates the source CEDAR template that was used to generate it; the linked template is later used to validate the user-created contents of the spreadsheet. These CEDAR-derived Excel constraints aim to discourage users from supplying metadata that does not conform to the specification in the source template. However, Excel does not force conformance to these constraints, so users can still supply nonconforming values.

Although spreadsheets provide great ease of use, it is simple for users to alter entries inadvertently and for the metadata to drift away from the standard. We thus developed an interactive Web-based application to ensure that metadata acquired via our CEDAR-generated spreadsheets adhere to source template specifications (18). Submitters use this application to interactively upload and then validate their spreadsheet-based metadata.

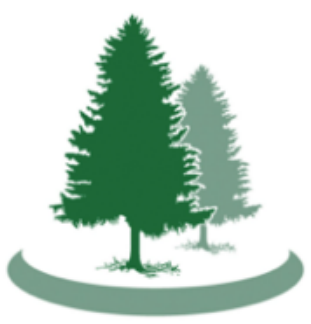

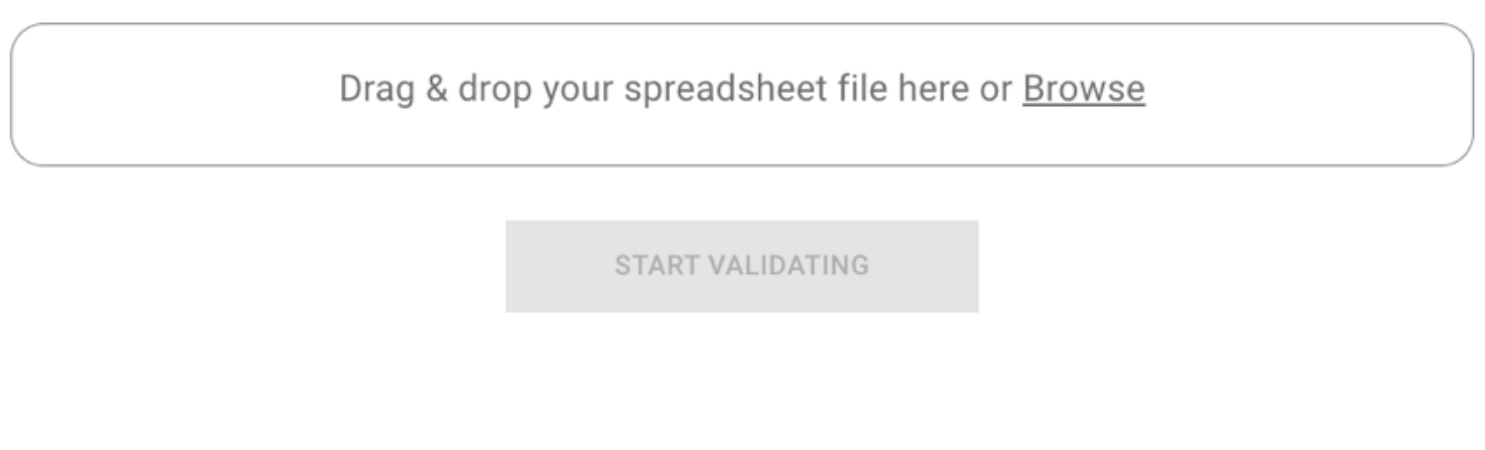

**Figure 5:** Landing page of CEDAR validator tool. Users can either drag-and-drop a metadata spreadsheet from their local computer to the input field or use the Browse option to select the spreadsheet from their file system. Once uploaded, a Start Validating option is enabled.

This tool adopts an array of strategies to ensure that the acquired metadata adhere to the source template specifications. It includes several wizard-style interfaces that focus first on reporting metadata errors in supplied spreadsheets and then on helping users to quickly repair those errors. When users upload a metadata spreadsheet to the tool (Figure 5), they are presented with a validation dashboard. This page displays a summary of the errors that were detected by the tool's algorithms (Figure 6).

The tool groups similar types of validation errors together and presents a high-level visual summary of these clusters so that users can quickly spot patterns in their errors. Two primary types of validation errors are addressed: (1) *completeness* errors and (2) *adherence* errors. Completeness validation concentrates on identifying required values that might be missing, whereas adherence validation aims to identify values that do not conform to the metadata specification (e.g., entries that are not elements of a required value set).

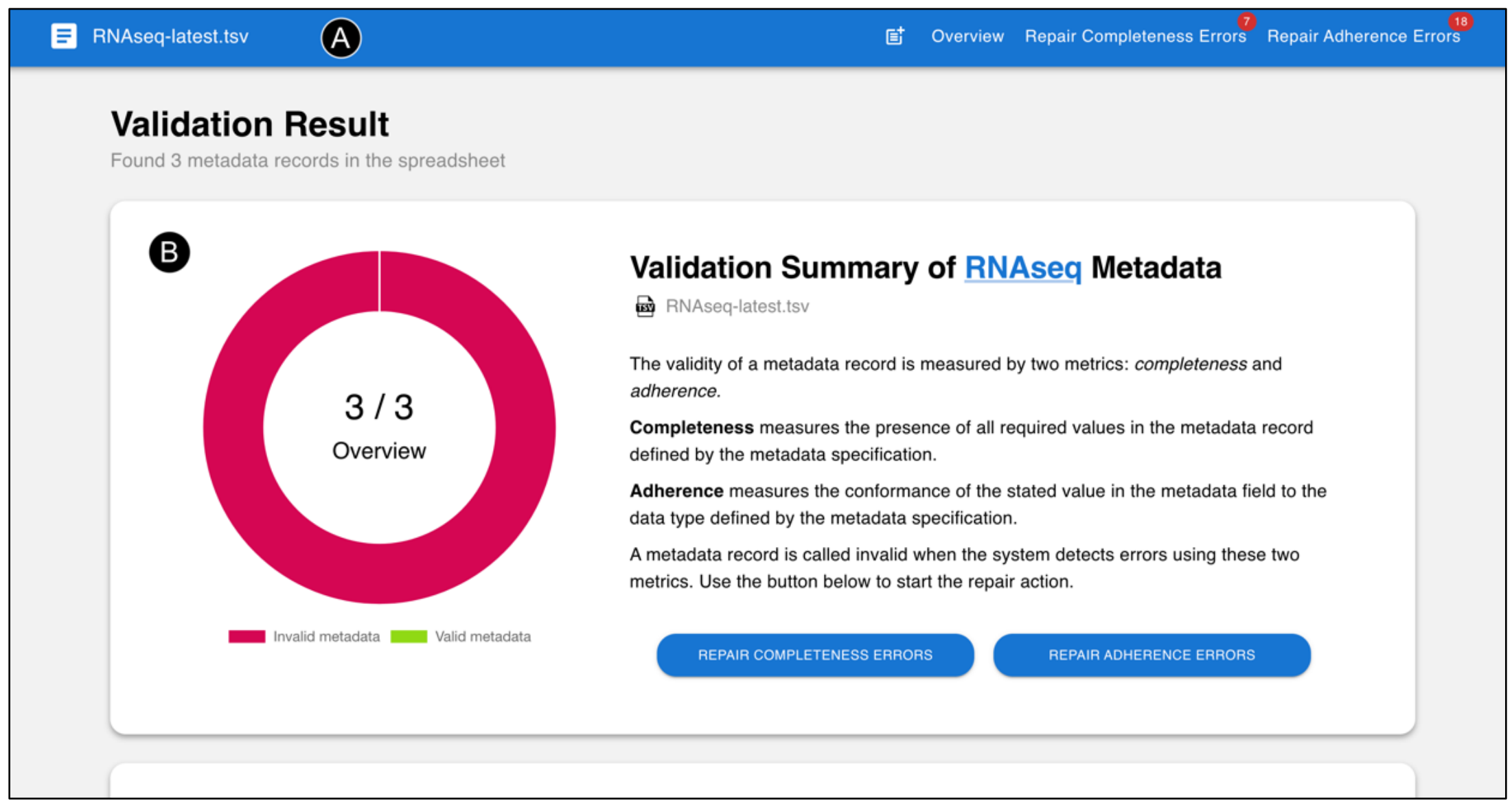

***Figure 6:*** CEDAR validator dashboard. The dashboard displays a summary of the errors detected in the uploaded metadata spreadsheet. These errors are divided into two types: completeness errors and adherence errors. Users can navigate through the errors using the top navigation bar (A). A donut chart (B) presents the total number of erroneous metadata records in the submitted spreadsheet.

To repair completeness errors (Figure 7), users are required to provide the missing values. This process involves presenting the identified information gaps to users and then assisting them in providing the appropriate input data to ensure the corresponding metadata record is complete and accurate.

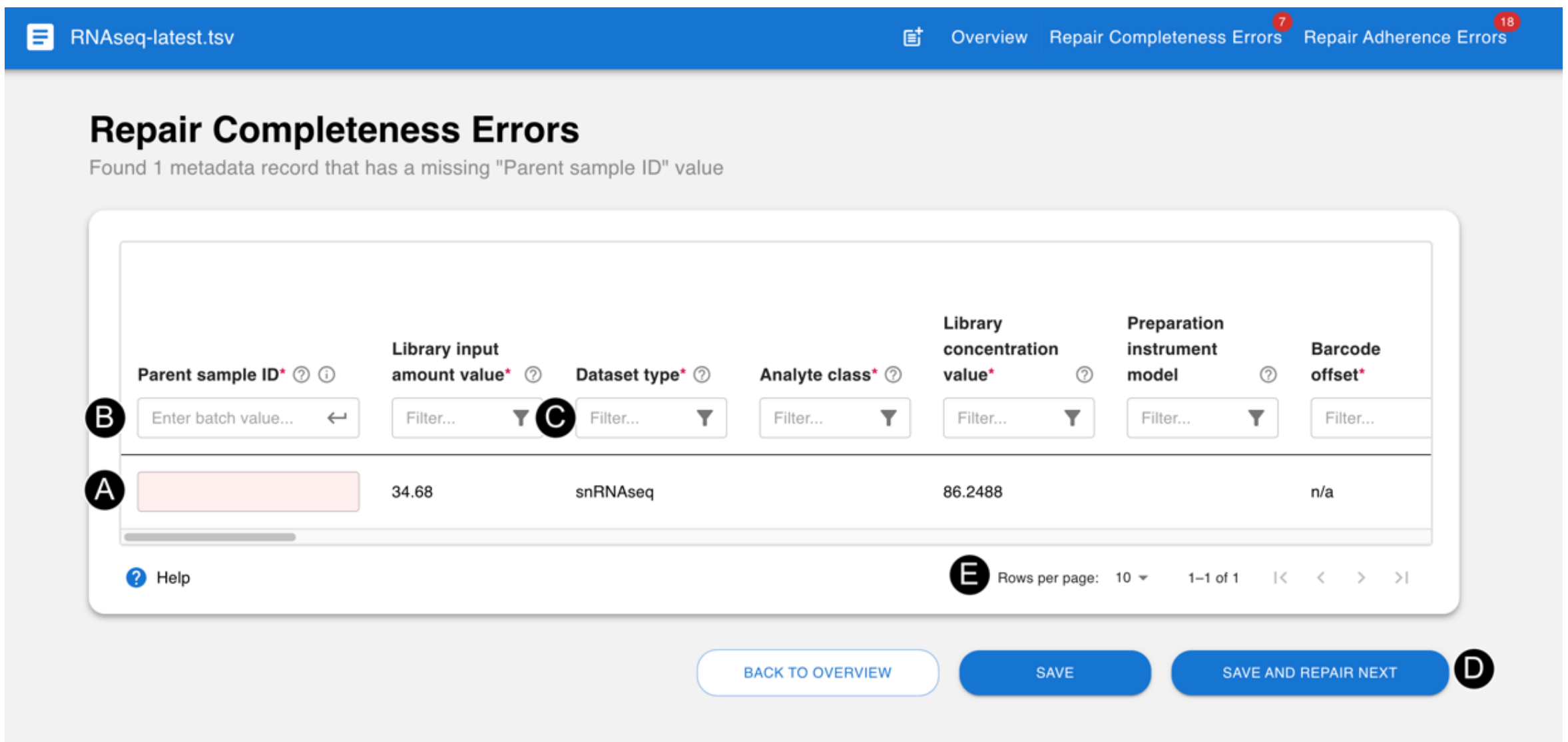

**Figure 7:** CEDAR validator's user interface for repairing completeness errors. Errors are paginated and presented in a tabular form. Fields in the left-most column (A) are used to enter the missing values. Batch repair of multiple rows for a column is also possible (B). In the case of batch repair, the value entered is applied to all displayed rows. Rows may also be filtered (C). At any point, users may save any changes (D) or navigate to additional pages of errors (E).

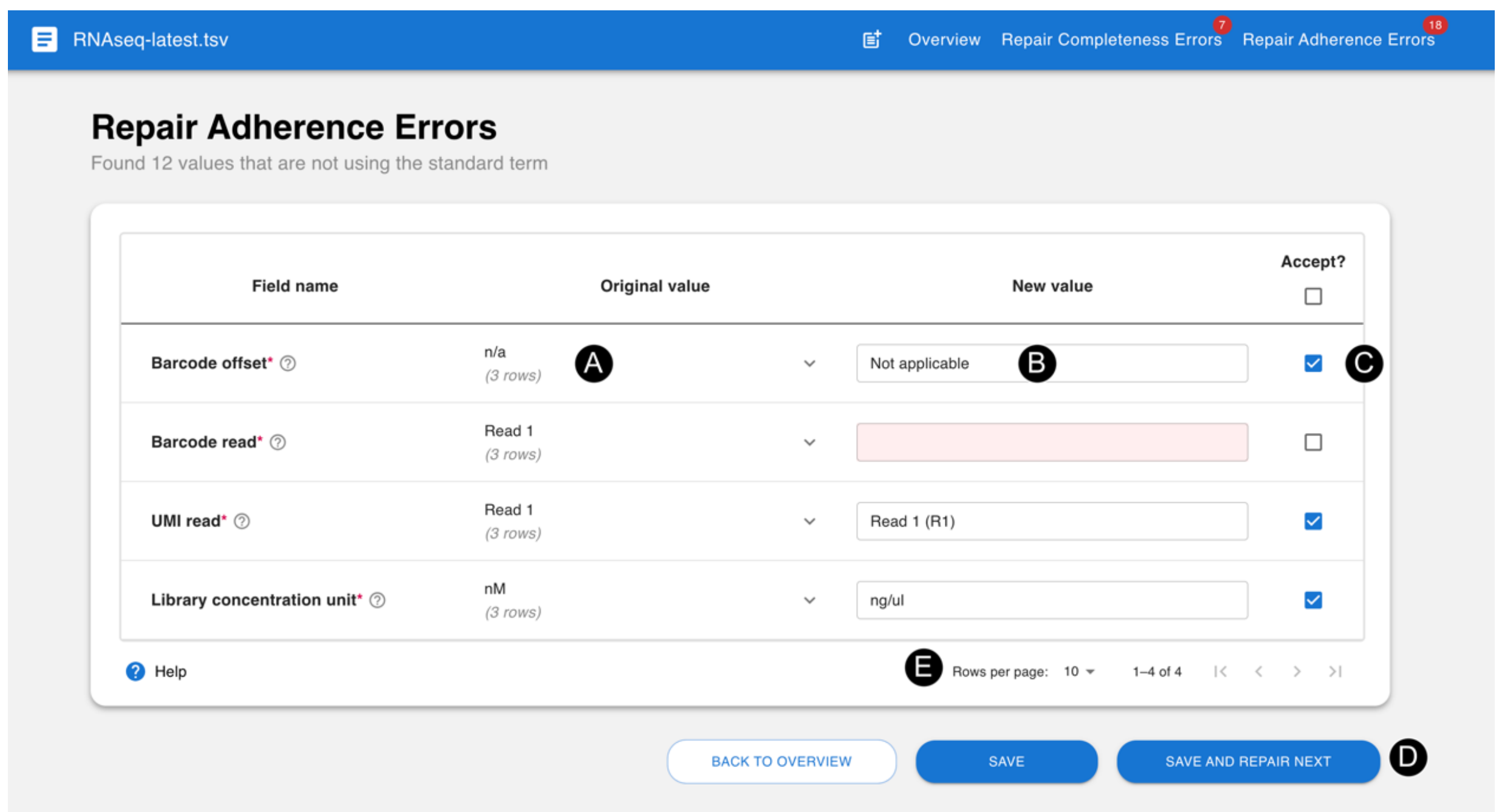

**Figure 8:** CEDAR validator's user interface for repairing adherence errors. Each row displays a column containing the values that appear to deviate from the standard (A). Users can enter the correct values directly or select from suggestions presented by the tool (B). Users can immediately accept the presented suggestions for a row (C). At any point, user may save the changes (D) or navigate to additional pages of errors (E).

A separate interface allows users to repair adherence errors (Figure 8). The tool repairs adherence errors semi-automatically by computing the most likely correct values and by presenting the resulting suggestions to the user, allowing them to quickly fix erroneous metadata. A key focus is assisting users in repairing adherence errors as effortlessly as possible by presenting suggestions for likely fixes. Simple suggestions may involve inferring likely values for supplied entries. For example, if the allowed values for a field are "Year", "Month", "Day" and a user supplies the value "days", the system can suggest "Day" as the corrected entry using a simple string distance metric. Other repairs supported by the validator include correcting simple typographical errors, such as, for example, removing quotation marks around numbers if the numbers were erroneously entered as string values.

More advanced approaches support metadata repair by exploiting the ability of CEDAR to specify ontology terms as allowed values for fields. Instead of simply looking at string-based values, the validator tool can use the template specification for that field to query the BioPortal ontology repository to get all possible values for a field together with synonyms of those values. Using this expanded list, the validator can effectively exploit the semantic information in field specifications. The validator tool uses this list infer the most likely similar value. The tool uses natural language processing capabilities in the large language model GPT 3.5 to select this value. As GPT 3.5 is trained on large tracts of text, it can establish similarities between pieces of text such as "formalin" and "methanol" without domain-specific training. GPT 3.5 is supplied with the user-provided metadata and all pertinent information (field name, field description) regarding the template field to enable it to make the choice between the possible value options.

By understanding the contextual meaning and relationships among these elements, the method can accurately identify and suggest the most relevant and contextually appropriate value, thus helping to ensure metadata integrity and consistency.

We developed a REST service that is accessed by our Web-based metadata validator to provide the necessary validation and repair functionality (19). This REST service is also used by HuBMAP's ingestion processes to validate all metadata uploaded to the system.

The entire workflow is driven from the CEDAR-based template that represents the relevant metadata reporting guideline, and no custom programming is needed. The CEDAR Workbench uses the relevant template to generate programmatically the spreadsheet that scientists use to enter their metadata. The metadata validator reads in the same template from CEDAR to check the spreadsheet-based metadata for errors. Once developers have represented the relevant reporting guideline in CEDAR, the rest of the metadata workflow falls into place automatically.

A shortcoming of the current implementation is that both the spreadsheet representation and the validation do not support multi-value fields. While this shortcoming did not prove relevant for HuBMAP metadata specifications, which are built from single-value fields only, it could prove problematic for standards that do contain such fields. We plan to address this issue in a future release. The approach will likely involve allowing multiple values to be inserted into an individual spreadsheet cell using a configurable separator character to delimit them. The validator will then be extended to handle such fields. Additional possibilities include developing a custom extension using Excel Office Ad-in that will support the intuitive editing and display of such fields within Excel.

We are also exploring the possibility of extending the validation approach to support metadata enrichment. This approach would involve leveraging a terminology service to augment entered free-text field values by associating them with precise identifiers from controlled vocabularies.

## Results

In the course of our project, all existing HuBMAP metadata reporting guidelines were replaced by machine-actionable metadata specifications defined using CEDAR templates. First, sample metadata specifications were replaced with CEDAR-based specifications. Then, we developed an array of CEDAR-based assay specifications for existing assay metadata reporting guidelines.

As each CEDAR-based metadata specification was finalized, we generated corresponding Excel-based representations. We also generated a human-readable representation of the specification. These Excel-based specifications together with the human-readable representation were then published on the HuBMAP Web site (20). The Excel files can be download by metadata submitters. Submitters can then populate these metadata spreadsheets. Once populated, they can validate these spreadsheets using the Web-based validation tool. Afterward, they can submit their metadata and associated raw data files to HuBMAP's data coordinating center.

**Table 1:** List of CEDAR-encoded HuBMAP metadata standards as of July 2024. All version 2.0.0 and later standards in HuBMAP use CEDAR's template-based mechanism for representing community-based standards.

| Specification Name | Version | Subject of Metadata |
|---|---|---|
| Sample Block | 2.1.0 | A solid piece of tissue extracted from a human donor |
| Sample Section | 2.1.0 | A slice of a sample block |
| Sample Suspension | 2.1.0 | Cells from a sample block in a liquid medium |
| Histology | 2.2.0 | Image of a tissue that shows microscopic features |
| Antibodies | 3.0.0 | Information about antibodies used in assays |
| Contributor | 2.0.0 | Individuals who contribute to the data set |
| GeoMx (NGS) | 2.0.0 | GeoMx next generation sequencing assays |
| GeoMx (nCounter) | 2.0.0 | GeoMx nCounter-based assays |
| Visium | 2.0.0 | Visium spatial gene expression assays |
| Visium (no probes) | 3.0.0 | Non-probe Visium spatial gene expression assays |
| Visium (with probes) | 3.0.0 | Probe-based Visium spatial gene expression assays |
| ATACseq | 3.0.0 | Assay for transposase-accessible chromatin |
| RNAseq (no probes) | 5.0.0 | Non-probe RNA sequencing assays |
| RNAseq (with probes) | 5.0.0 | Probe-based RNA sequencing assays |
| 10X Multiome | 2.0.0 | Multi-assay that combines ATACseq and RNAseq |
| SnareSeq2 | 2.0.0 | Single-nucleus chromatin accessibility and RNAseq assays |
| HiFi-Slide | 2.0.0 | Slide-based high-fidelity assays |
| MIBI | 2.0.0 | Multiplexed ion beam imaging assays |
| IMC 2D | 2.0.0 | 2D image mass cytometry assays |
| LC-MS | 4.0.0 | Liquid chromatography mass spectrometry assays |
| NanoSplits | 2.0.0 | Nanodroplet splitting for linked-multimodal investigations of trace samples assays |
| DESI | 2.0.0 | Desorption electrospray ionization assays |
| MALDI | 2.0.0 | Matrix-assisted laser desorption/ionization assays |
| SIMS | 2.0.0 | Secondary ion mass spectrometry assays |
| CODEX | 2.0.0 | Co-detection by indexing assays |
| Auto-flourescence | 2.1.0 | Autofluorescence microscopy assays |
| Cell DIVE | 2.1.0 | Cellular dynamics in visual environment assays |
| Light Sheet | 3.1.0 | Light sheet fluorescence microscopy assays |
| Confocal | 2.1.0 | Confocal microscopy assays |
| CyCIF | 2.1.0 | Cyclic immunofluorescence assays |
| Enhanced SRS | 2.1.0 | Enhanced stimulated raman scattering assays |
| PhenoCycler | 2.2.0 | PhenoCycler multiplex fluorescence microscopy assays |
| SHG | 2.1.0 | Second harmonic generation assays |
| Thick Section Multiphoton MxIF | 2.1.0 | Thick section multiphoton multiplexed immunofluorescence assays |
| MUSIC | 2.0.0 | Multiplexed single-cell in situ cytometry assays |
| MERFISH | 2.0.0 | Multiplexed error-robust fluorescence in situ hybridization assays |

The existing HuBMAP submission processes were modified to handle these Excel-based specifications. Since there is no guarantee that submitters have pre-validated their metadata using the Web-based validation tool, the HuBMAP submission system re-validates the metadata using our REST-based validation interfaces. The ingestion and validation of these metadata within the HuBMAP system are driven by the CEDAR templates that represent the relevant standards. Thus, all spreadsheet-encoded metadata uploaded by users are validated against their relevant CEDAR template in a dynamic fashion. If validation fails, a detailed report is returned to the invoking service indicating the reason for the failure. This report is presented to metadata submitters in a user-friendly format. Submitters must repair the errors and resubmit their dataset. All currently submitted HuBMAP datasets have used our REST-based metadata-validation service. HuBMAP put the CEDAR-based metadata workflow into production in August 2023. Currently, there are 34 CEDAR-encoded HuBMAP metadata standards in production use (Table 1).

Several new metadata reporting guidelines are under development and will be released over the next year.

Many of the current 34 standards have gone through multiple versioned releases. In general, these updates were driven by minor additions or modifications to individual metadata fields. HuBMAP used CEDAR's template versioning infrastructure to handle these updates. When a major change to a standard was needed (for example, renaming a field or adding a new required field), a new CEDAR template was released with the corresponding new version number, replacing the earlier representation of the standard. However, for minor, backwards-compatible revisions, we regenerated and reuploaded the Excel files to the HuBMAP Web site without changing the version number. If a metadata specification update entailed changes in the controlled terms used for field values, we updated the value set and submitted the updated version of the value set to BioPortal, effectively replacing the previous version.

Beyond supporting downstream dataset ingestion, processing, and publication, the CEDAR-based validation in HuBMAP provides an opportunity to help educate data submitters on metadata best practices. For instance, HuBMAP's central data curators use the Web-based validator tool as an instructional aid during the project's Data Submission Office Hours. In these sessions, a central curator often onboards personnel from data-provider teams who will be uploading data and metadata to the consortium's data portal. As part of this training, the curator demonstrates the usage of the Web-based validator and explains how the submitter can use the tool to check their metadata before formally submitting to HuBMAP.

The central HuBMAP curation team has found that the Web-based metadata validator has dramatically simplified their metadata repair workflow. Data submitters now engage with metadata validation errors more proactively than reactively. Previously, when a central curator identified a validation issue, the ingestion process would need to be halted. The curator and the data submitter would then need to troubleshoot the metadata, with the central curator needing to identify and recommend fixes to be made. Currently, data providers engage with central curation much earlier in the data submission process, thanks to the Web-based validator. For example, submitters will immediately notice when metadata elements with categorical values fail validation. Frequently these errors are related to new instrumentation or reagent kits used in their sequencing experiments. When this type of error occurs, submitters reach out to the central curation team, who then coordinate with internal HuBMAP project managers and the CEDAR developers via a standardized workflow. With this approach, value sets and assay metadata templates are then updated, usually within the same business day, empowering efficient, standards-adherent metadata record creation.

## Discussion

CEDAR technologies provide support for metadata specification, acquisition, validation, and repair capabilities in the HuBMAP project. Through the collaborative development environment offered by CEDAR, we created high-quality metadata templates for describing HuBMAP sample and assay metadata. CEDAR then automatically converts these specifications into spreadsheets, which dataset submitters use to upload their multi-assay metadata to HuBMAP. An interactive Web-based service provided by CEDAR enabled users to validate and repair metadata supplied using these spreadsheets, ensuring that the acquired metadata meet the quality specifications provided in the source templates. These capabilities offer a robust, end-to-end metadata management solution that addresses all stages of the metadata lifecycle, enabling users to quickly fix erroneous metadata and submit a repaired spreadsheet along with associated data. Overall, the use of CEDAR technologies has greatly enhanced the metadata management capabilities in the HuBMAP project.

The FAIR Guiding Principles have proven challenging to operationalize. Investigators and policy makers alike have run into difficulty determining when these principles have been successfully implemented. This difficulty stems from the subjective nature of many of the principles. For example, the FAIR Guiding Principles call for metadata that are “rich” and that are “adherent to community standards.” But what makes metadata “rich”, and how does one know which community standard to follow (if any even exists) in any given situation? When attempting to create FAIR datasets in scientific communities, the challenges for creating valid metadata become pressing. The FAIR principles provide general guidelines, but they provide little practical assistance during the detailed work of creating actual metadata descriptions. Clear definitions of what constitutes FAIR-adherent metadata need to come from scientific communities themselves. Unless the metadata preferences for a scientific community can be encoded, preserved, communicated, and enforced in a consistent manner, the promise of FAIR datasets will continue to remain elusive (13).

In many scientific communities, the goal of creating high-quality metadata to describe scientific datasets predates the introduction of the FAIR Guiding Principles by several decades. These efforts focus on the creation of community standards that define the necessary metadata to describe scientific experiments. A large number of such standards have been developed in recent decades in a variety of scientific domains. Again, however, these efforts suffer from shortcomings similar to those of the FAIR Guiding Principles. Central to these shortcomings is a lack of precision when specifying individual metadata field values. A metadata reporting guideline may, for example, indicate that a metadata field should specify a disease, but may be silent on how the value of that field should be encoded.

As outlined in this paper, CEDAR aims to address the limitations of both the FAIR Guiding Principles and community-based metadata reporting guidelines, focusing on the development of detailed, machine-readable templates to produce FAIR metadata. In the work that we present in this paper, we propose that communities of investigators should create machine-processable metadata templates that represent their own relevant standards and that guide data stewards in how those standards should be applied. These metadata templates provide canonical representations of the reporting guidelines and associated ontologies important to a given community. Such templates serve as a cornerstone for capturing the preferences of a scientific community regarding standardized metadata practices. These templates allow investigators to describe all the "data about the data" necessary to understand the nature of a study, its motivation, and the methods used. A filled-out metadata template provides a comprehensive, machine-readable summary of everything needed to interpret the study and assess the reusability of the data. These templates enable both humans and computers to access these representations within a tool ecosystem that enhances data FAIRness. By specifying metadata guidelines in advance of dataset production and supporting the evaluation and correction of existing metadata according to community standards, these templates play a crucial role in ensuring that data are FAIR.

Once community-based metadata standards are represented as templates, developers can then create tools that use the knowledge in those templates; these templates can effectively facilitate plug and play with arbitrary software systems. Our template model provides a straightforward mechanism to translate the knowledge of textual reporting guidelines into a machine-actionable format. Because a metadata template can be easily read and processed by a variety of applications (mitigating vendor lock-in and associated blocks to interoperability), it can form the basis for a standard, technology-independent means to communicate metadata reporting guidelines in a computable fashion. Our intention is not to introduce yet another redundant standard to make an already complex landscape of standards even more confusing. Rather, we aim to provide a mechanism for representing an existing type of standard (namely, metadata reporting guidelines) more precisely and in a way that is more readily actionable. There is currently no agreed-upon convention for how reporting guidelines should be represented, and the availability of a coherent format that is compatible with widely used knowledge-representation standards provides an obvious advantage.

The work outlined in this paper demonstrates the utility of a standard format and illustrates how tools can be developed that use it to provide a comprehensive end-to-end metadata workflow for a particular scientific community. A key focus of this work is supporting interoperation with spreadsheets. Spreadsheets are the *lingua franca* of data and metadata exchange in many scientific communities, and any approach to creating good metadata must support interoperation with these simple tables. Other communities may have other standard formats. The beauty of our approach is that you create a template once in CEDAR, and then it is available for any number of tools.

The spreadsheet-generation process and the metadata-validation tools described in this paper are generic and are not HuBMAP-specific. There are, however, some limitations to the current tools. Most of these limitations reflect the rigidity of spreadsheets, which, for example, lack the ability to incorporate branching logic for acquisition of cell values. Future developments may focus on enhancing the flexibility and sophistication of spreadsheet-based metadata entry systems to address such limitations. We also plan to perform a comprehensive evaluation of the quality of the metadata collected during the HuBMAP deployment. Currently, only a portion of the expected final number of HuBMAP datasets have been submitted. The vast majority of submissions are expected over the next year or two, after which we will perform a summative evaluation of the quality of all metadata.

While the solution outlined here is tailored to meet the needs of a specific project, we believe its applicability extends to any metadata-acquisition scenario where interoperation with existing data repositories is required. The adaptability of CEDAR templates supports the development of tools that allow seamless integration with other data repositories, facilitating the generalization of our approach. The representation of community-based metadata standards as templates in CEDAR effectively allows for transfer of knowledge regarding metadata preferences across software systems. A key characteristic of the technologies described in this paper is the ability to use them out-of-the-box to meet the needs of other scientific communities. For example, CEDAR templates are currently being used to interoperate with a variety of other data formats used by the Dryad (21), Open Science Framework (22), and HEAL (23) data repositories. Using these capabilities, customized templates can be built to meet the specific metadata reporting guidelines of different research initiatives, enabling researchers to capture the necessary information to understand study details, motivation, and methods. Once developed, an array of existing CEDAR tools can then be used to work with these templates. In particular, the spreadsheet creation and validation capabilities outlined in this paper are immediately available for communities that wish to utilize a spreadsheet-based metadata workflow. A workflow using these technologies is already in use by the Cellular Senescence Network (SenNet) program to describe the metadata in their system (24). We believe that the tools provided by CEDAR can be similarly adopted to serve the metadata management needs of a wide array of groups.

Metadata templates are a fundamental mechanism by which groups of investigators can capture and communicate their requirements for the metadata needed to make datasets FAIR. Our template model offers a straightforward mechanism for translating textual reporting guidelines into a machine-actionable format. Because the model can be easily processed by various applications, it forms the basis for a standard, technology-independent means to communicate reporting guidelines. Our approach demonstrates that a formal metadata model can standardize reporting guidelines and enable software systems to assist in the authoring of standards-adherent metadata. The inherent rigor, precision, and reusability of machine-actionable metadata templates support these activities, ultimately leading to better data and better science.

Our findings demonstrate the effectiveness of metadata templates in representing community-based metadata standards and ensuring adherence to those standards, which can improve data sharing and reuse in scientific research. The work also shows that, when a scientific community's preferences regarding metadata standards can be encoded in a representation such as that used by CEDAR, the computer-stored representation can be deployed in a range of contexts, serving as a reference for that can inform a variety of software systems that each address different elements of a data-management ecosystem.

## Acknowledgments

This work was supported in part by award OT2 OD033759 from the U.S. National Institutes of Health Common Fund, by grant R01 LM013498 from the U.S. National Library of Medicine, and by grant U24 GM143402 from the U.S. National Institute of General Medical Sciences. We are grateful to HubMAP Consortium members Phil Blood, Bill Shirey, Jonathan Silverstein, Alan Simmons and Richard Morgan for helpful discussions about this work.

## Data Availability

HuBMAP datasets are accessible through the HuBMAP Portal (see https://portal.hubmapconsortium.org). HuBMAP metadata specifications can be reviewed at https://hubmapconsortium.github.io/ingest-validation-tools/current.

## Code Availability

The software described in this paper is available for execution at https://metadatavalidator.metadatacenter.org, https://cedar.metadatacenter.org and https://hubmapconsortium.github.io/ingest-validation-tools. All source code is available from the Metadata Center landing page at https://github.com/metadatacenter and the HuBMAP landing page at https://github.com/hubmapconsortium.

# Author Contributions

## Authors and Affiliations

**Stanford Center for Biomedical Informatics Research, Stanford University, Stanford, CA 94305, USA**
Martin J. O'Connor, Josef Hardi, Marcos Martínez-Romero, Sowmya Somasundaram and Mark A. Musen

**Pittsburgh Supercomputer Center, Pittsburgh, PA 15213, USA**
Brendan Honick

**University of Pennsylvania**
Stephen A. Fisher

**National Human Genome Research Institute, Bethesda, MD 208792, USA**
Ajay Pillai

### Contributions

M.A.M. and A.P. conceived the concept. J.H., M.J.O'C., M.M.R. and S.S. developed the spreadsheet export and validation software. J.H. and S.A.F. were the main contributors to the development the HuBMAP metadata templates outlined in this paper. B.H. provided curation support for HuBMAP metadata collection. M.J.O'C. wrote the initial draft of the manuscript. All authors reviewed and edited the manuscript. M.J.O'C and J.H. contributed equally to this work.

### Corresponding Author

Martin J. O'Connor

## Ethics Declarations

### Competing Interests

The authors declare no competing interests.